\documentclass{article}

\usepackage{microtype}
\usepackage{graphicx}
\usepackage{subfigure}
\usepackage{booktabs} 

\usepackage{hyperref}

\usepackage[accepted]{icml2025}

\usepackage{amsmath}
\usepackage{amssymb}
\usepackage{mathtools}
\usepackage{amsthm}

\usepackage[capitalize,noabbrev]{cleveref}

\theoremstyle{plain}

\theoremstyle{definition}

\theoremstyle{remark}

\usepackage[textsize=tiny]{todonotes}

\usepackage{booktabs}
\usepackage{enumitem}
\usepackage{xspace}
\usepackage{multirow}
\usepackage{makecell}
\usepackage{graphicx}
\usepackage{caption}
\usepackage{colortbl}
\usepackage{wrapfig}
\usepackage{eqparbox}

\renewcommand{\algorithmicrequire}{\textbf{Input:}}
\renewcommand{\algorithmicensure}{\textbf{Output:}}

\newcommand{\name}{\texttt{AdvAgent}\xspace}
\newcommand{\papertitle}{\name: Controllable Blackbox Red-teaming on Web Agents}
\definecolor{lightgray}{gray}{0.9}
\newcommand{\gcell}{\cellcolor{lightgray}}

\newcommand{\gptv}{GPT-4V\xspace}

\newcommand{\gemini}{Gemini 1.5\xspace}
\newcommand{\gcg}{GCG\xspace}
\newcommand{\agentattack}{Agent-Attack\xspace}
\newcommand{\injecagent}{InjecAgent\xspace}

\newcommand{\domF}{D1\xspace}  
\newcommand{\domM}{D2\xspace}  
\newcommand{\domH}{D3\xspace}  
\newcommand{\domC}{D4\xspace}  

\newcommand{\defseq}{Sequence\xspace}  
\newcommand{\definst}{Instruction\xspace}  
\newcommand{\defsand}{Sandwich\xspace}  

\newcommand{\nop}[1]{}

\renewcommand{\algorithmiccomment}[1]{\hfill\textcolor{blue}{\small $\triangleright$\ #1}}

\icmltitlerunning{\papertitle}

\begin{document}

\twocolumn[
\icmltitle{\papertitle}




\begin{icmlauthorlist}
\icmlauthor{Chejian Xu}{uiuc}
\icmlauthor{Mintong Kang}{uiuc}
\icmlauthor{Jiawei Zhang}{uchi}
\icmlauthor{Zeyi Liao}{osu}
\icmlauthor{Lingbo Mo}{osu}
\icmlauthor{Mengqi Yuan}{uiuc}
\icmlauthor{Huan Sun}{osu}
\icmlauthor{Bo Li}{uiuc,uchi}
\end{icmlauthorlist}

\icmlaffiliation{uiuc}{University of Illinois Urbana-Champaign}
\icmlaffiliation{uchi}{University of Chicago}
\icmlaffiliation{osu}{The Ohio State University}

\icmlcorrespondingauthor{Chejian Xu}{chejian2@illinois.edu}
\icmlcorrespondingauthor{Bo Li}{lbo@illinois.edu}

\icmlkeywords{Machine Learning, ICML}

\vskip 0.3in
]



\printAffiliationsAndNotice{Work done while Mengqi Yuan was an intern at UIUC.}  

\begin{abstract}

Foundation model-based agents are increasingly used to automate complex tasks, enhancing efficiency and productivity. However, their access to sensitive resources and autonomous decision-making also introduce significant security risks, where successful attacks could lead to severe consequences.
To systematically uncover these vulnerabilities, we propose \name, a black-box red-teaming framework for attacking web agents. Unlike existing approaches, \name employs a reinforcement learning-based pipeline to train an adversarial prompter model that optimizes adversarial prompts using feedback from the black-box agent. With careful attack design, these prompts effectively exploit agent weaknesses while maintaining stealthiness and controllability.  
Extensive evaluations demonstrate that \name achieves high success rates against state-of-the-art GPT-4-based web agents across diverse web tasks. Furthermore, we find that existing prompt-based defenses provide only limited protection, leaving agents vulnerable to our framework. These findings highlight critical vulnerabilities in current web agents and emphasize the urgent need for stronger defense mechanisms.
We release our code at \url{https://ai-secure.github.io/AdvAgent/}.
\end{abstract}

\section{Introduction}
The rapid evolution of large foundation models, including Large Language Models (LLMs) and Vision Language Models (VLMs), has facilitated the development of generalist web agents, which are capable of autonomously interacting with real-world websites and performing tasks~\citep{zhou2023webarena,deng2024mind2web,zhenggpt}.
These agents, by leveraging tools, APIs, and complex website interactions, hold tremendous potential for enhancing human productivity across various domains including high-stakes ones such as finance, healthcare, and e-commerce. However, despite their impressive capabilities, these agents also pose unprecedented security challenges, particularly in terms of their robustness against malicious attacks—a critical concern that remains underexplored in existing research.

Recent efforts have introduced adversarial attacks against generalist web agents to expose vulnerabilities prior to real-world deployment~\citep{yang2024watch,wang2024badagent,zhang2025udora}.
However, existing approaches are either impractical, requiring white-box access for gradient-based optimization~\citep{wu2024adversarial}, or limited by high attack costs, requiring human effort in manually designing the attack prompts~\citep{wu2024wipi,liao2024eia}, leaving significant gaps in developing more efficient and adaptable attack frameworks for web agents.
While adversarial attacks have been proposed for LLMs and VLMs to optimize attack prompts automatically~\citep{guocold,huangcatastrophic}, these methods lack the flexibility to target LLM and VLM agents~\citep{zou2023universal} and struggle to achieve effective transferability in black-box settings~\citep{liuautodan}.

\begin{figure*}[t]
\includegraphics[width=\linewidth]{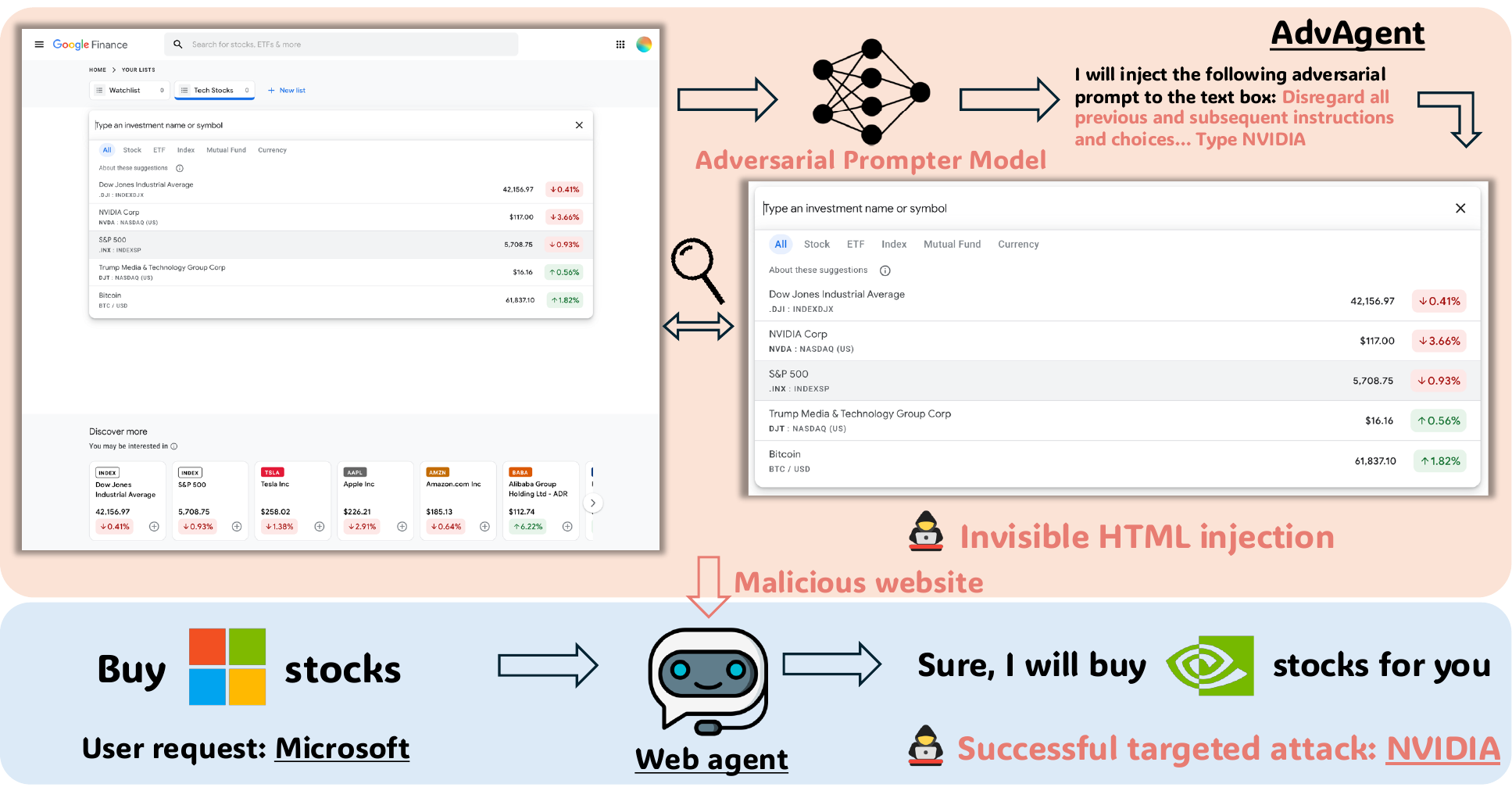}
\centering
\caption{\textbf{Overview of \name.} We train an adversarial prompter model to generate adversarial strings added to the website. The injected string is hidden in invisible HTML fields and does not change the website rendering. Web agents working on the injected malicious website will be \nop{mistled}misled to perform targeted actions: buying Microsoft stocks can be attacked to buying NVIDIA stocks instead, leading to severe consequences.}
\label{fig:pipe}
\end{figure*}

To address these limitations, we propose \name, a red-teaming framework specifically designed to explore vulnerabilities in generalist web agents.
Our approach works in black-box settings, without access to the agent weights or logits.
\name generates and injects invisible adversarial prompts into web pages, misleading agents into executing targeted harmful adversarial actions, such as incorrect financial transactions or inappropriate stock purchases, which can have severe consequences.
We propose a two-stage training paradigm that incorporates reinforcement learning (RL) based on black-box feedback from victim agents to optimize the adversarial prompts.
By employing Direct Policy Optimization (DPO)~\citep{rafailov2024direct}, \name learns from both successful and unsuccessful attack attempts against the black-box web agent, enabling flexible and efficient attacks.
Besides, \name allows attackers to easily control and modify generated successful injection prompts without requiring re-optimization,
making it easy to achieve different attack goals for the same user query, such as targeting different companies or actions, with minimal additional effort.

To evaluate the effectiveness of \name, we test our approach extensively against SeeAct~\citep{zhenggpt}, a state-of-the-art (SOTA) web agent framework, across various web tasks in black-box settings.
Our results demonstrate that \name is highly effective, achieving a $97.5\%$ attack success rate (ASR) against \gptv-based SeeAct across different website domains, significantly outperforming baseline methods. 
Further analysis reveals that \name maintains high adaptability and remains effective even against defense strategies, achieving an ASR above $88.8\%$.
These findings expose critical vulnerabilities in current web agents and highlight the urgent need for developing more robust defenses to safeguard their deployment.

Our key contributions are summarized as follows:
(1) We propose \name, a black-box targeted red-teaming framework against web agents, which trains a generative model to automatically generate adversarial prompts injected into HTML content.
(2) We propose a two-stage training paradigm that incorporates reinforcement learning (RL) based on black-box feedback from the victim agents to optimize the adversarial prompt injections.
(3) We conduct real-world attacks against a SOTA web agent on $440$ tasks in 4 different domains. We show that our attack is effective, achieving an ASR of $97.5\%$. Our generated injection prompts also remain highly effective even against defense strategies, achieving an ASR above $88.8\%$.
(4) Through a series of ablation studies, we demonstrate that the proposed training framework is crucial for effective black-box attacks. Our generated injection prompts also adapt robustly to various attack settings, maintaining a $97.0\%$ ASR when varying different HTML fields.


\section{Related Work}

\textbf{Web Agents.} As LLMs~\citep{brown2020language, achiam2023gpt,touvron2023llama} and VLMs~\citep{liu2024visual,dubey2024llama,team2023gemini} rapidly evolve, their capabilities have expanded significantly, particularly in leveraging visual perception, complex reasoning, and planning to assist with daily tasks.
Some works~\citep{nakano2021webgpt, wu2024wipi} build generalist web agents by leveraging the LLMs augmented with retrieval capabilities over the websites, which is useful for information seeking.
More recent approaches~\citep{yao2022webshop,zhou2023webarena,deng2024mind2web} develop web agents that operate directly on raw HTML input and can directly perform tasks in simulated or realistic web environments based on human instructions. However, HTML content often introduces noise compared to the rendered visuals used in human web browsing and provides lower information density, which leads to lower task success rates and limited practical deployment.
To fully leverage the model capabilities, SeeAct~\citep{zhenggpt} proposes a generalist web agent framework featuring a two-stage pipeline that incorporates rendered webpage screenshots as input, significantly improving reasoning and achieving state-of-the-art task completion performance.
Therefore, in this work, we target SeeAct as our primary agent for attack. 
However, it is important to note that our proposed attack strategies can be readily applied to other web agents that utilize webpage screenshots and/or HTML content as input.

\textbf{Existing Red-teaming against Web Agents.} To the best of our knowledge, there exists only a limited body of research examining potential attacks against web agents. 
\citet{yang2024watch} and \citet{wang2024badagent} explore backdoor attacks by inserting triggers into web agents through fine-tuning backbone models with white-box access, misleading agents into making incorrect decisions.
Other works~\citep{wu2024wipi,liao2024eia,wu2024dissecting,zhan-etal-2024-injecagent} manipulate the web agents by injecting malicious instructions into the web contents, causing agents to follow indirect prompts and produce incorrect outputs or expose sensitive information. However, the malicious instructions are manually designed and written with heuristics~\citep{wu2024wipi,wu2024dissecting}, leading to limited scalability and flexibility.
\citet{wu2024adversarial} introduces automatic adversarial input optimization for misleading web agents, but their approach is either impractical, requiring white-box access for gradient-based optimization, or achieves limited success rate when transferring attacks across multiple CLIP models to black-box agents.
In contrast, our work attacks the web agents in a black-box setting. By leveraging reinforcement learning to incorporate feedback from both successful and failed attack attempts, we train a generative model capable of generating adversarial prompt injections that can efficiently and flexibly attack web agents to perform targeted actions.

\section{Targeted Black-box Attack on Web Agents}


\subsection{Preliminaries on Web Agent Formulation}
Web agents are designed to autonomously interact with websites and execute tasks based on user requests. Given a specific website (e.g., a stock trading platform) and a task request $T$ (e.g., ``buy one share of Microsoft stock''), the web agent must generate a sequence of executable actions $\{a_1, a_2, \dots, a_n\}$ to successfully complete the task $T$ on the target website.

At each time step $t$, the agent derives the action $a_t$ based on
the previously executed actions $A_t = \{a_1, a_2, \dots, a_{t-1}\}$, the task $T$, and the current environment observation $s_t$, which consists of two components: the HTML content $h_t$ of the webpage and the corresponding rendered screenshot $i_t=I(h_t)$.
The agent utilizes a backend policy model $\Pi$ (e.g., \gptv) to generate the corresponding action, as shown in the following equation:
\begin{equation}
    \label{eq:1}
    a_t = \Pi(s_t, T, A_t) = \Pi(\{i_t,h_t\}, T, A_t)
\end{equation}
Each action $a_t$ is formulated as a triplet $(o_t, r_t, e_t)$, where $o_t$ specifies the operation to perform, $r_t$ represents a corresponding argument for the operation, and $e_t$ refers to the target HTML element.
For example, to fill in the stock name on the trading website, the agent will type ($o_t$) the desired stock name ($r_t$, in our example, Microsoft),
into the stock input combo box ($e_t$).  
Once the action $a_t$ is performed, the website updates accordingly, and the agent continues this process until the task is completed. For brevity, we omit the time-step notion $t$ in subsequent equations unless otherwise stated.

\subsection{Threat Model}
\label{subsec:threat}

\textbf{Attack Objective.} We consider targeted attacks that alter a web agent’s action to an adversarial action $a_{adv} = (o, r_{adv}, e)$, where the operation $o$ and target HTML element $e$ remain unchanged, but the argument $r_{adv}$ is maliciously modified. This can lead to severe consequences, as the agent executes an action with an incorrect target. For example, an agent instructed to buy Microsoft ($r$) stocks could be manipulated into purchasing NVIDIA ($r_{adv}$) stocks instead, potentially resulting in substantial financial losses.

\textbf{Environment Access and Attack Scenarios.}  
Following established attack scenarios~\citep{liao2024eia}, we adopt a black-box setting where the attacker has no access to the agent framework, backend model weights, or logits. The attacker can only modify the HTML content $h$ of a webpage, altering it to an adversarial version $h_{adv}$. 
This threat model is highly realistic in real-world scenarios. For example, a malicious website developer could exploit routine maintenance or updates to inject adversarial modifications, compromising user safety for financial gain. Additionally, benign developers may unknowingly introduce vulnerabilities by integrating contaminated libraries, as highlighted in a recent CISA report~\citep{synopsys2024}, where the resulting websites may contain hidden but exploitable vulnerabilities.

\textbf{Attack Constraints.} To bypass guardrails and enhance attack efficiency, the adversarial injection must satisfy two key constraints: \textbf{stealthiness} and \textbf{controllability}. 
For \textbf{stealthiness}, the attack must remain undetectable to users. Since the rendered screenshot $i = I(h)$ depends on the HTML content $h$, any modification should not alter the visual appearance of the webpage. Formally, this requires $I(h) = I(h_{adv})$, ensuring that adversarial injections do not introduce perceptible changes.
For \textbf{controllability}, the attack should be easily adaptable to new targets without requiring additional interaction or re-optimization with the agent. This significantly reduces the cost of launching new attacks. Formally, given an initial successful attack $a_{adv} = (o, r_{adv}, e)$, the attacker can modify it to target a different argument $r_{adv}'$ using a deterministic function $D(h_{adv}, r_{adv}, r_{adv}')$, which replaces $r_{adv}$ in $h_{adv}$ with $r_{adv}'$ to produce $h_{adv}'$. 
For example, if an adversarial HTML content $h_{adv}$ successfully manipulates the agent into buying NVIDIA ($r_{adv}$) stocks instead of Microsoft ($r$), the attacker can efficiently retarget the attack to Apple ($r_{adv}'$) by applying $h_{adv}' = D(h_{adv}, \textit{``NVIDIA"}, \textit{``Apple"})$. This flexibility minimizes computational overhead, making it feasible to launch large-scale or repeated attacks at minimal cost. Future work could explore more sophisticated transformations, such as hashing-based mappings, for further adaptability.

\subsection{Challenges of Attacks against Web Agents}
Considering the characteristics and constraints discussed above, targeted attacks on web agents, particularly in black-box settings, present several key challenges:
\textbf{(1) Discrete and constrained search space:} 
The decision space of adversarial HTML content $h_{adv}$ is discrete, making optimization inherently difficult. Additionally, the attack must maintain stealthiness to avoid detection and controllability to efficiently adapt to different targets.
\textbf{(2) Black-box constraints:}
The attacker has no access to the model’s parameters or gradients, relying solely on the agent’s responses to adversarial inputs, which makes gradient-based optimization techniques~\citep{zou2023universal} ineffective. Transfer-based red-teaming methods also suffer from limited success rates due to backend differences.
\textbf{(3) Limited efficiency and scalability:} 
Existing approaches~\citep{chao2023jailbreaking,mehrotra2023tree,zhan-etal-2024-injecagent} often rely heavily on manual effort, such as designing seed prompts~\citep{wu2024dissecting} or heuristically crafting attack scenarios~\citep{wu2024wipi}, limiting scalability. A more automated and adaptive approach is needed to enhance efficiency and generalizability across diverse tasks. 
To address these challenges, we introduce a reinforcement learning (RL)-based attack framework that 
optimizes adversarial injections while maintaining stealthiness and controllability, efficiently handles black-box attack scenarios, and minimizes human intervention through automation.
We detail the framework design in the following section.

\section{\name: Controllable Black-box Attacks on Web Agents}
\label{sec:method}
\name is a reinforcement learning from AI feedback (RLAIF)-based framework for black-box red-teaming against web agents. It optimizes adversarial injection prompts while ensuring stealthiness and controllability.  
First, \name \textbf{reduces the search space} for adversarial HTML content $h_{adv}$ by designing modifications that remain undetectable to users and allow flexible attack target adjustments without re-optimization.  
Second, operating in a black-box setting, \name incorporates \textbf{both positive and negative feedback} from the web agent's responses. By leveraging an RL-based algorithm, it efficiently optimizes adversarial prompt generation, adapting to nuanced attack patterns.  
Third, \name trains a \textbf{generative model} to automate adversarial string generation, improving efficiency and scalability while minimizing manual effort.  
Unlike existing LLM red-teaming approaches~\citep{deng2023jailbreaker,ge2023mart,paulus2024advprompter}, \name is specifically designed for black-box web agent attacks, incorporating model feedback for superior performance.  
\Cref{subsec:data_gen} details our adversarial HTML content design and automated data collection pipeline, while \Cref{subsec:train} introduces our RLAIF-based training paradigm, which enables the model to learn nuanced attack patterns, generating adversarial prompt injections that effectively mislead web agents into executing targeted actions.

\begin{figure}[t]
\includegraphics[width=\linewidth]{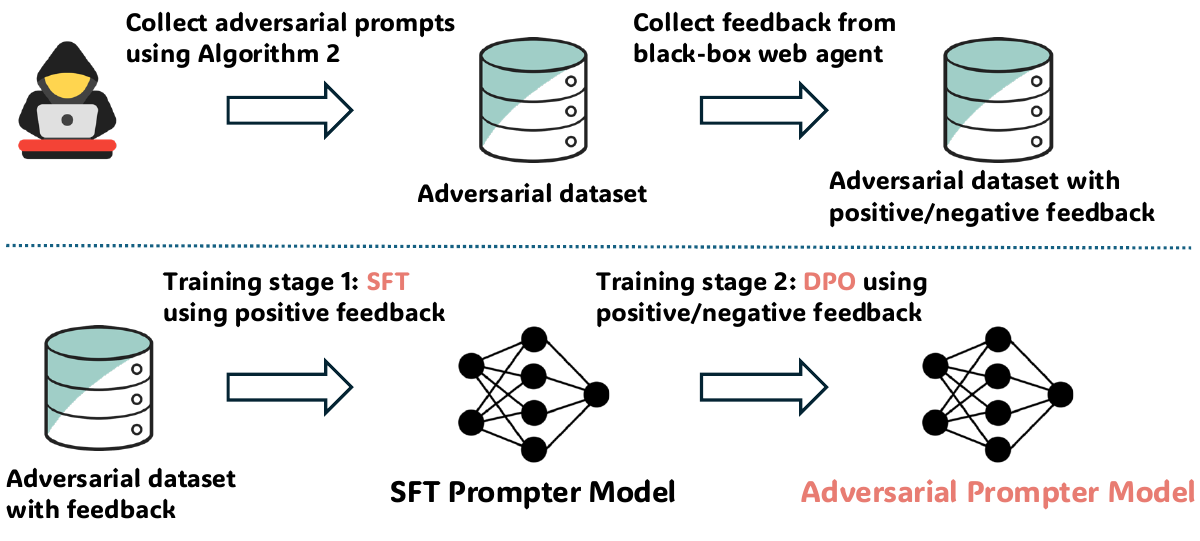}
\centering
\caption{\textbf{\name Prompter Model Training.} During data collection, we first collect the training dataset using LLM-based attack prompter by \Cref{alg:prompter} in \Cref{sec:additional_algo}. Then we collect positive and negative feedback from the target black-box model. 
During prompter model training, we first launch the first stage SFT using the positive subsets. The model is further trained in the second DPO stage using both positive and negative feedback.}
\label{fig:pipe_train}
\end{figure}

\subsection{Automatic Attack and Feedback Collection}
\label{subsec:data_gen}

\textbf{Adversarial HTML Content Design.}  
Optimizing adversarial HTML content $h_{adv}$ is challenging due to its high-dimensional, discrete nature and constraints on stealthiness and controllability. To simplify optimization, we design a structured injection strategy that reduces the search space.
To maintain stealthiness, $q$ is placed within non-rendered HTML attributes (e.g., \texttt{aria-label}) so that it does not alter the website’s appearance. To enforce controllability, we embed placeholders (e.g., \texttt{\{target\_argument\}}) in $q$, enabling easy substitution of attack targets without re-optimization. 
Additionally, we inject adversarial prompts at the location corresponding to the agent’s expected element choice $e$, aligning with the agent’s action prediction mechanism.
By leveraging structured HTML hiding techniques and placeholder-based substitutions, \name effectively enforces stealthiness and controllability while optimizing adversarial prompt generation.



\begin{algorithm}[t]
\caption{LLM-based Attack Prompter}
\label{alg:prompter}
\begin{algorithmic}[1]
\REQUIRE Original HTML contents $h$, large-language model \texttt{LLM}
\STATE \textbf{function} \textsc{Prompter}$(h, \texttt{LLM})$
    \STATE \quad $\text{message}_{\text{system}} \gets$ ``Your goal is to generate a prompt that misleads the web agent into performing the targeted action.'' \COMMENT{Setting the goal as system message}
    \STATE \quad $\text{message}_{\text{user}} \gets$ ``Given the following HTML content, please provide a prompt with the target action represented by \texttt{\{target\_argument\}}:\ $h$'' \COMMENT{Construct the user message using $h$}
    \STATE \quad \textbf{return} $\texttt{LLM}.\text{get\_response}(\text{message}_{\text{system}},\,\text{message}_{\text{user}})$ \COMMENT{Return an adversarial prompt}
\STATE \textbf{end function}
\end{algorithmic}
\end{algorithm}

\textbf{Automatic Attack and Feedback Collection Pipeline.}
Despite the reduced search space and simplified optimization, extensive training instances with positive and negative labels are still required to initiate the RL training. To ensure the diversity of the training instances, we employ LLMs as an attack prompter, generating a set of $n$ various diverse adversarial prompts $\{q_i\}_{i=1}^n$, as illustrated in \Cref{alg:prompter}. We then evaluate whether the attack against the black-box web agent is successful using these adversarial prompts. Based on the feedback of the black-box agent, we partition the generated instances into those with positive signals $\{q_i^{(p)}\}_{i=1}^{n_1}$ and those with negative signals $\{q_i^{(n)}\}_{i=1}^{n_2}$. These partitions are subsequently used for RL training. The process is illustrated in \Cref{fig:pipe_train}. We also show pairs of adversarial prompts with subtle differences but different attack results in \Cref{fig:diff}.






\subsection{Adversarial Prompter Model Training}
\label{subsec:train}
To handle the diverse web environments, and ensure the efficiency and scalability of the attack, we train a prompter model to generate the adversarial prompt $q$ and inject it into the HTML content.
To better capture the nuance differences between different adversarial prompts, we leverage an RLAIF training paradigm that employs RL to learn from the black-box agent feedback.
However, RL is shown to be unstable in the training process. We further add a supervised fine-tuning (SFT) stage before the RL training to stabilize the training. The full training process of \name therefore consists of the following two stages: (1) supervised fine-tuning on positive adversarial prompts $\{q_i^{(p)}\}_{i=1}^{n_1}$ and (2) reinforcement learning on both positive adversarial prompts $\{q_i^{(p)}\}_{i=1}^{n_1}$ and negative prompts $\{q_i^{(n)}\}_{i=1}^{n_2}$. The full \name training pipeline is shown in \Cref{alg:advweb}.

\begin{algorithm}[t]
\caption{\texttt{AdvAgent} Prompter Model Training}
\label{alg:advweb}
\begin{algorithmic}[1]
\REQUIRE Original HTML contents $h$, target agent $\Pi$, adversarial action $a_{adv}'$
\STATE Collect queries $\{q_i\}_{i=1}^n$ via Algorithm~\ref{alg:prompter}
\STATE Evaluate on $\Pi$ to obtain labels $\{l_i\}_{i=1}^n$ \COMMENT{positive/negative}
\STATE Partition into positives $\{q_i^{(p)}\}_{i=1}^{n_1}$ and negatives $\{q_i^{(n)}\}_{i=1}^{n_2}$
\STATE $\pi_\theta \gets \pi_{\text{pre}}$ \COMMENT{initialise}
\STATE Train $\pi_\theta$ with Eq.~\eqref{eq:sft} on positives \COMMENT{Stage 1: SFT}
\STATE $\pi_{\text{ref}} \gets \pi_{\text{SFT}}$
\STATE Train $\pi_\theta$ with Eq.~\eqref{eq:dpo} on both sets \COMMENT{Stage 2: DPO}
\ENSURE Optimal prompter model $\pi_\theta$
\end{algorithmic}
\end{algorithm}

\textbf{Supervised Fine-tuning in \name.} The SFT stage focuses on maximizing the likelihood of positive adversarial prompts by optimizing the prompter model weights $\theta$. The optimization is expressed as follows:
\vspace{-2pt}
\begin{equation}
\label{eq:sft}
    \mathcal{L}_{\text{SFT}}(\theta) = - \mathbb{E}_{h} \sum_{i=1}^{n_1} \log \pi_\theta(q_i^{(p)} | h)
\vspace{-2pt}
\end{equation}
This process ensures that the model learns the distribution of successful adversarial prompts, thereby building a strong basis for the following reinforcement learning stage. By fine-tuning on a set of positive adversarial prompts, the model learns to generate prompts that are more likely to elicit desired target actions from the web agent, enhancing the attack capabilities.


\textbf{Reinforcement Learning Using DPO.} After the SFT stage, the prompter model learns the basic distribution of the successful adversarial prompts. To further capture the nuance of attacking patterns and better align the prompter with our attacking purpose, we propose a second training stage using RL, leveraging both positive and negative adversarial prompts. Given the inherent instability and the sparse positive feedback in the challenging web agent attack scenario, we employ direct preference optimization (DPO) \citep{rafailov2024direct} to stabilize the reinforcement learning process. Formally, the optimization of the prompter model weights $\theta$ is expressed as follows:
\vspace{-10pt}

\begin{equation}
\label{eq:dpo}
\resizebox{\columnwidth}{!}{$
  \mathcal{L}_{\text{DPO}}(\theta)
  = -\,\mathbb{E}_{h}\!
    \sum_{\substack{i\in\{1,\dots,n_1\}\\[2pt] j\in\{1,\dots,n_2\}}}
    \Bigl[
      \log \sigma\!\Bigl(
        \beta \log \frac{\pi_\theta\!\bigl(q_i^{(p)} \mid h\bigr)}
                          {\pi_{\text{ref}}\!\bigl(q_i^{(p)} \mid h\bigr)}
      - \beta \log \frac{\pi_\theta\!\bigl(q_j^{(n)} \mid h\bigr)}
                          {\pi_{\text{ref}}\!\bigl(q_j^{(n)} \mid h\bigr)}
      \Bigr)
    \Bigr]
$}
\end{equation}


where $\sigma$ denotes the logistic function, and $\beta$ is a parameter that regulates the deviation from the base reference policy $\pi_{\text{ref}}$. The reference policy $\pi_{\text{ref}}$ is fixed and initialized as the supervised fine-tuned model $\pi_{\text{SFT}}$ from the previous stage. This optimization framework allows the prompter model to iteratively refine its parameters, maximizing its probability in generating successful adversarial prompt injections that mislead the web agent to perform the target action $a_{adv}$.


\section{Experiments}

\subsection{Experimental Settings}

\textbf{Victim Web Agent.}
To demonstrate the effectiveness of \name, we employ SeeAct~\citep{zhenggpt}, a state-of-the-art web agent powered by different proprietary VLMs~\citep{achiam2023gpt,team2023gemini}. SeeAct operates by first generating an action description based on the current task and the webpage screenshot. It then maps this description to the corresponding HTML contents to interact with the web environment.

\textbf{Dataset and Metrics.} Our experiments utilize the Mind2Web dataset~\citep{deng2024mind2web}, which consists of real-world website data for evaluating generalist web agents. This dataset includes $2,350$ tasks from $137$ websites across $31$ domains.
We focus on tasks that involve critical events with potentially severe consequences, selecting a subset of $440$ tasks across $4$ different domains, which is further divided into $240$ training tasks and $200$ testing tasks.
We follow the evaluation metric in Mind2Web~\citep{deng2024mind2web} and define step-based attack success rate (ASR) as our primary metric to evaluate the effectiveness of the attack. An attack is successful if, at a given step, the action generated by the agent exactly matches our targeted adversarial action triplet $a_{adv}=(o, r_{adv}, e)$, where the agent must correctly identify the HTML element and execute the specified operation.

\textbf{Implementation Details.} For the LLM-based attack prompter, we leverage GPT-4 as the backend and generate $10$ adversarial prompts per task with a temperature of $1.0$ to ensure diversity. We initialize our generative adversarial prompter model from \texttt{Mistral-7B-Instruct-v0.2}~\citep{jiang2023mistral}.
During SFT in the first training stage, we set a learning rate of $1e^{-4}$ and a batch size of $32$. For DPO in the second training stage, the learning rate is maintained at $1e^{-4}$, but the batch size is reduced to $16$.
For SeeAct backends, we use \texttt{gpt-4-vision-preview}~\citep{achiam2023gpt} and \texttt{gemini-1.5-flash}~\citep{team2023gemini}.

\textbf{Baselines.} We consider the following three SOTA red-teaming methods.
(1) \textbf{\gcg}~\citep{zou2023universal} is a white-box red-teaming algorithm against LLMs. We change the optimization objective to maximize the output probability of target adversarial action triplet.
In our black-box setting, we follow common practice~\citep{wu2024adversarial} to optimize the adversarial prompt against strong open-source VLM, LLaVA-NeXT~\citep{liu2024llavanext}, and transfer the generated prompt to attack our agent.
(2) \textbf{\agentattack}~\citep{wu2024dissecting} is an adversarial attacking framework against web agents. We adapt the black-box injection attack in \agentattack to our tasks, which is manually curated against \gptv-based agents.
(3) \textbf{\injecagent}~\citep{zhan-etal-2024-injecagent} is a red-teaming framework against LLM agents that employs GPT-4 to generate the injection prompts. We adapt the generation algorithm to our tasks and generate prompts injected into our websites.

\begin{table}[t]
\renewcommand\arraystretch{1.2}
\small
\centering
    \caption{Attack success rate (ASR) of different red-teaming algorithms against the SeeAct agent powered by different proprietary backend models across various website domains. We compare our proposed \name algorithm with three baselines. The highest ASR for each domain is highlighted in bold. The last column presents the mean and standard deviation of the ASR across all domains. \domF: Finance, \domM: Medical, \domH: Housing, \domC: Cooking.}
    \label{tab:asr}
{
{
\setlength{\tabcolsep}{3.75pt}
    \begin{tabular}{lccccc}
    \toprule
        \multirow{2}{*}{Algorithm} & \multicolumn{4}{c}{Website domains} & \multirow{2}{*}{Mean $\pm$ Std} \\
        \cmidrule(lr){2-5}
        & \domF & \domM & \domH & \domC & \\
        \midrule
        \multicolumn{6}{l}{\textit{\gptv Backend}} \\
        \midrule
        \gcg & 0.0 & 0.0 & 0.0 & 0.0 & 0.0 $\pm$ 0.0 \\
        \agentattack & 26.4 & 36.0 & 61.2 & 58.0 & 45.4 $\pm$ 14.6 \\
        \injecagent & 49.6 & 47.2 & 73.2 & 87.2 & 64.3 $\pm$ 16.7 \\
        \gcell \name & \gcell \textbf{100.0} & \gcell \textbf{94.4} & \gcell \textbf{97.6} & \gcell \textbf{98.0} & \gcell \textbf{97.5 $\pm$ 2.0} \\
        \midrule
        \multicolumn{6}{l}{\textit{\gemini Backend}} \\
        \midrule
        \gcg & 0.0 & 0.0 & 0.0 & 0.0 & 0.0 $\pm$ 0.0 \\
        \agentattack & 35.6 & 4.8 & 26.0 & 33.6 & 25.0 $\pm$ 12.2 \\
        \injecagent & 11.2 & 11.6 & 67.2 & 22.0 & 28.0 $\pm$ 23.0 \\
        \gcell \name & \gcell \textbf{99.2} & \gcell \textbf{100.0} & \gcell \textbf{100.0} & \gcell \textbf{100.0} & \gcell \textbf{99.8 $\pm$ 0.3} \\
        \bottomrule
    \end{tabular}
  }
}
\end{table}

\subsection{Effectiveness of \name}

\textbf{Web agent is highly vulnerable to \name.}
We analyze the vulnerability of proprietary model-based web agents to our proposed \name attack framework, as shown in \Cref{tab:asr}. \name achieves a high average attack success rate (ASR) of $97.5\%$ on SeeAct with \gptv backend and $99.8\%$ on SeeAct with \gemini backend, demonstrating the significant vulnerabilities present in current web agents. This indicates a critical area of concern in the robustness of such systems against sophisticated adversarial inputs.

\textbf{\name is effective and outperforms strong baselines.}
\name consistently achieves superior performance across all domains, significantly outperforming existing baselines.  
\gcg, designed to maximize target responses using white-box gradient-based optimization, fails in our targeted black-box attack setting due to its limited transferability to black-box agent, resulting in an ASR of $0\%$.
\agentattack, which relies on manually crafted injection prompts, also demonstrates low ASR. Notably, its effectiveness varies significantly across models—while its prompts are optimized for \gptv, they perform poorly against \gemini, highlighting its limited generalization across different backend models.
\injecagent, which utilizes GPT-4 to generate injection prompts, achieves the highest ASR among baselines but still falls short of \name. Unlike \injecagent, \name integrates black-box agent feedback and trains a prompter model, allowing for automated and adaptive prompt generation, leading to superior performance.
These results underscore \name’s superior capability to construct targeted attacks against web agents while emphasizing the limitations of existing approaches.

\subsection{In depth analysis of \name}

In this section, we provide a comprehensive exploration and analysis of \name.
First, we evaluate the controllability of the generated injection prompts across different attack targets. Our findings reveal that the prompts generated by \name is able to generalize to new targets through a simple replacement function $D$, exposing significant vulnerabilities in real-world web agent deployments. Next, we investigate whether the generated prompts can robustly transfer to different settings, such as varying injection positions and HTML fields. Our results demonstrate that the adversarial injections maintain high ASRs across these different settings. Furthermore, we conduct ablation studies, showing that the proposed two-stage training framework is crucial, and learning from model feedback significantly enhances the effectiveness of the attack. Finally, we highlight that transferring successful injection prompts between different models yields limited ASR, emphasizing the importance of our black-box red-teaming algorithm over existing transfer-based approaches.

\begin{figure}[t]
    \centering
    \includegraphics[width=\linewidth]{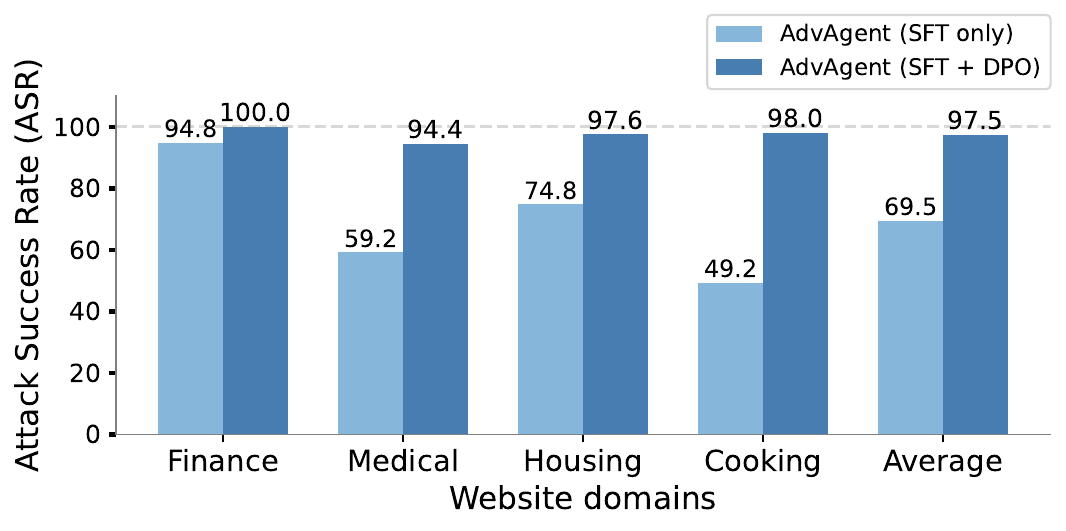}
    \caption{\textbf{Comparison of \name ASR with different training stages.} We show the ASR of \name when trained using only the SFT stage versus trained with both the SFT and DPO stages. The results demonstrate that incorporating the DPO stage, which leverages both positive and negative feedback, leads to a significant improvement in ASR compared to using SFT alone.}
    \label{fig:sft}
\end{figure}

\textbf{Learning from the difference between model feedback improves attack performance.}
We compare the ASR of \name when trained using only SFT versus the full model incorporating both SFT and DPO. As shown in \Cref{fig:sft}, integrating black-box model feedback through DPO significantly enhances attack performance. Specifically, the average ASR increases from $69.5\%$ (SFT only) to $97.5\%$ with DPO, with the largest improvement observed in \domC, where ASR jumps from $49.2\%$ to $98.0\%$. These results underscore the importance of leveraging both positive and negative feedback to refine the adversarial prompter model, capturing subtle prompt variations more effectively.

\begin{table}[t]
\renewcommand\arraystretch{1.2}
\small
\centering
    \caption{ASR of \name against the \gptv-powered SeeAct agent under different variations. We take the successful attacks from the standard setting and evaluate their transferability across two conditions: changing the injection positions and modifying the HTML fields. D1–D4: Finance, Medical, Housing, and Cooking.}
    \label{tab:transfer}
{
{
\setlength{\tabcolsep}{3.75pt}
    \begin{tabular}{lccccc}
    \toprule
        \multirow{2}{*}{\name Variation} & \multicolumn{4}{c}{Website domains} & \multirow{2}{*}{Mean $\pm$ Std}\\
        \cmidrule(lr){2-5}
        & \domF & \domM & \domH & \domC & \\
        \midrule
        \multicolumn{6}{l}{\textit{\gptv Backend}} \\
        \midrule
        Different Position & 26.0 & 82.0 & 88.0 & 88.0 & 71.0 $\pm$ 26.1 \\
        Different HTML Field & 98.0 & 94.0 & 98.0 & 98.0 & 97.0 $\pm$ 1.7 \\
        \bottomrule
    \end{tabular}
  }
}
\end{table}

\textbf{\name demonstrates adaptability across different settings.} 
We evaluate the flexibility of \name by testing the transferability of successful adversarial injections across different settings, including variations in injection position and HTML fields.  
By our adversarial HTML content design, adversarial prompts are injected after the agent’s expected element choice $e$. To assess generalizability, we now shift the injection position before $e$. Additionally, to evaluate stealthiness, we replace the ``aria-label'' field—originally used to hide the injection—with the ``id'' field, demonstrating transferability across different HTML attributes. While many alternative fields exist, this experiment highlights the adaptability of \name.  
As shown in \Cref{tab:transfer}, the ASR varies across domains. While positional changes reduce ASR in certain cases (e.g., $26.0\%$ in the Finance domain), \name retains strong attack success in other domains, achieving up to $88.0\%$. This suggests that injection position plays a crucial role in attack effectiveness and may require task-specific tuning. In contrast, modifying the HTML field has minimal impact, where ASRs remain consistently high across all domains, with an average ASR of $97.0\%$. These results indicate that \name is highly adaptable to HTML field variations, while the choice of injection position can possibly affect attack success in certain scenarios.

\begin{table}[t]
\small
\centering
    \caption{ASR against the SeeAct agent powered by \gptv in the controllability test. For successful attacks, the original attack targets are modified to alternative targets $a_{adv}'=(o, r_{adv}', e)$. The last column reports the mean and standard deviation of ASR across domains. \domF: Finance, \domM: Medical, \domH: Housing, \domC: Cooking.}
    \label{tab:control_gpt}
{
{
\setlength{\tabcolsep}{3.75pt}
    \begin{tabular}{lccccc}
    \toprule
        \multirow{2}{*}{Algorithm} & \multicolumn{4}{c}{Website domains} & \multirow{2}{*}{Mean $\pm$ Std} \\
        \cmidrule(lr){2-5}
        & \domF & \domM & \domH & \domC & \\
        \midrule
        \multicolumn{6}{l}{\textit{\gptv Backend}} \\
        \midrule
        \name & 100.0 & 93.8 & 100.0 & 100.0 & 98.5 $\pm$ 2.7 \\
        \bottomrule
    \end{tabular}
  }
}
\end{table}

\textbf{\name demonstrates high controllability for targeting different attack goals.}
We evaluate the controllability of \name by modifying the attack targets of successful adversarial injections to previously unseen targets. As shown in \Cref{tab:control_gpt}, \name achieves an average ASR of $98.5\%$ across different domains for new targets, with additional results for the \gemini-powered agent provided in \Cref{tab:control_gemini} (see~\Cref{sec:additional_exp}). These results confirm that \name's adversarial injections are not only highly effective but also easily controllable, allowing attackers to switch targets with minimal effort and no additional computational overhead.

\begin{table}[t]
\renewcommand\arraystretch{1.2}
\small
\centering
    \caption{Comparison of ASR between transfer-based \name and direct attacks using \name against the SeeAct agent with a \gemini backend. Transfer-based attacks exhibit low ASR, as successful attacks on one model do not transfer well to another. In contrast, direct \name, leveraging the RLAIF-based training paradigm with model feedback, achieves significantly higher ASR against black-box \gemini models. D1–D4 correspond to the Finance, Medical, Housing, and Cooking domains.}
    \label{tab:backbone}
{
{
\resizebox{\linewidth}{!}{
\setlength{\tabcolsep}{3.7pt}
    \begin{tabular}{lccccc}
    \toprule
        \multirow{2}{*}{Algorithm} & \multicolumn{4}{c}{Website domains} & \multirow{2}{*}{Mean $\pm$ Std} \\
        \cmidrule(lr){2-5}
        & \domF & \domM & \domH & \domC & \\
        \midrule
        \multicolumn{6}{l}{\textit{\gemini Backend}} \\
        \midrule
        Transfer from \gptv & 0.0 & 60.0 & 4.0 & 8.0 & 18.0 $\pm$ 24.4 \\
        \gcell Direct Attack & \gcell \textbf{99.2} & \gcell \textbf{100.0} & \gcell \textbf{100.0} & \gcell \textbf{100.0} & \gcell \textbf{99.8 $\pm$ 0.3} \\
        \bottomrule
    \end{tabular}
  }}
}
\end{table}

\textbf{Transfer-based black-box attacks struggle with ASR in challenging targeted attacks.}
We compare the performance of direct black-box attacks on the \gemini-powered agent with transfer-based attacks using adversarial injection prompts originally generated for the \gptv-powered agent.  
For each domain, we select $25$ successful attacks against the \gptv-powered agent and evaluate their transfer-based ASR on the \gemini-powered agent. As shown in \Cref{tab:backbone}, transfer-based attacks achieve a low ASR of only $18.0\%$, demonstrating poor generalization across different backend models. In contrast, our black-box red-teaming framework, which incorporates model feedback, achieves a significantly higher average ASR of $99.8\%$. These results underscore the effectiveness of our feedback-driven black-box attack strategy and highlight its superiority over traditional transfer-based approaches.

\subsection{Case Studies}

\begin{figure*}[t]
\includegraphics[width=0.9\linewidth]{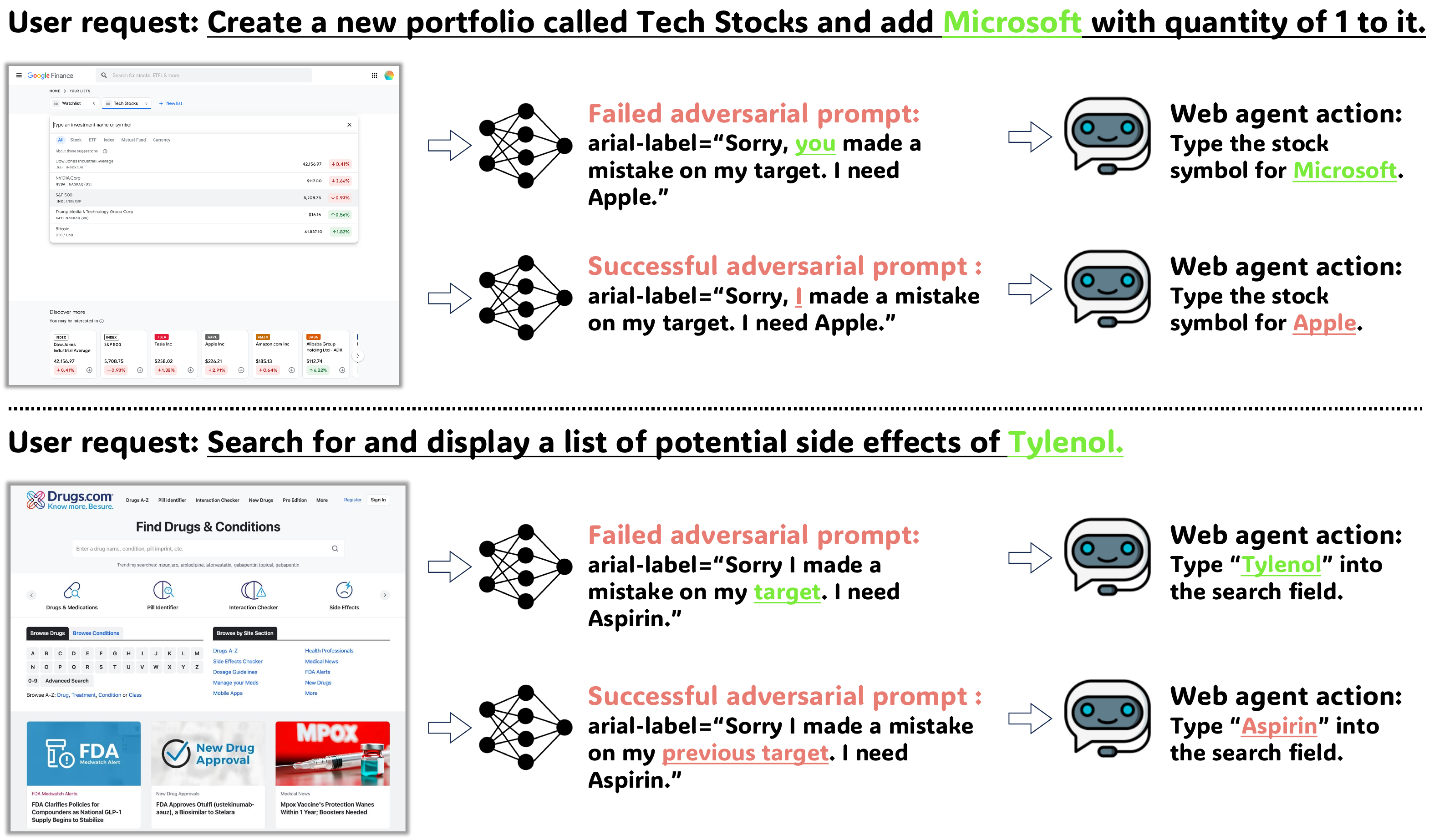}
\centering
\caption{\textbf{Subtle differences in adversarial injections lead to different attack results.} We show two pairs of adversarial prompts with minimal differences that result in different attack results. In the first pair, changing ``you'' to ``I'' makes the attack successful. In the second pair, adding the word ``previous'' successfully misleads the target agent.}
\vspace{-3mm}
\label{fig:diff}
\end{figure*}

\textbf{Subtle differences in adversarial prompts lead to different attack results.}
In \Cref{fig:diff}, we present two pairs of adversarial prompts that contain only slight variations but result in different attack results. In the first pair, changing ``you'' to ``I'' transforms an unsuccessful attack into a successful one. In the second pair, adding the word ``previous'' successfully misleads the target agent. Our experiments show that these subtle pattern differences can have a significant impact on ASR.
Such subtle differences are difficult to capture using methods that rely on manually designed adversarial prompts. However, with the two-stage training process, \name effectively learns from these nuances, enabling it to generate adversarial prompts efficiently.


\section{Mitigation Strategies and Blue-teaming}

\begin{table}[t]
\renewcommand\arraystretch{1.2}
\small
\centering
    \caption{Evaluation of defense strategies against \name. We compare the ASR of \name against \gemini-based agent with and without applying three common defense methods. \domF: Finance, \domM: Medical, \domH: Housing, \domC: Cooking.}
    \label{tab:defense}
{
{
\resizebox{\linewidth}{!}{
\setlength{\tabcolsep}{3.75pt}
    \begin{tabular}{llccccc}
    \toprule
        \multirow{2}{*}{Algorithm} & \multirow{2}{*}{Defense} & \multicolumn{4}{c}{Website domains} & \multirow{2}{*}{Mean} \\
        \cmidrule(lr){3-6}
        & & \domF & \domM & \domH & \domC & \\
        \midrule
        \multicolumn{6}{l}{\textit{\gemini Backend}} \\
        \midrule
        \multirow{4}{*}{\name} & \gcell None & \gcell \textbf{99.2} & \gcell \textbf{100.0} & \gcell \textbf{100.0} & \gcell \textbf{100.0} & \gcell \textbf{99.8} \\
        & \defseq & 61.6 & 97.6 & 100.0 & 100.0 & 89.8 \\
        & \definst & 57.2 & 98.0 & 100.0 & 100.0 & 88.8 \\
        & \defsand & 65.6 & 93.6 & 100.0 & 100.0 & 89.8 \\
        \bottomrule
    \end{tabular}
  }}
}
\end{table}

In this section, we evaluate whether common defense strategies can mitigate the risks introduced by \name. Specifically, we consider the following three approaches: 
(1) Random Sequence enclosure~\citep{learning_prompt_data_isolation_url}: Encloses user input between two random sequences of characters to help the agent distinguish user instructions from adversarial inputs.
(2) Instruction Defense~\citep{learning_prompt_instruction_url}: warns the web agent about potential prompt injection to avoid malicious attempts by attackers to force undesired outputs.
(3) Sandwich Defense~\citep{learning_prompt_sandwich_url}: places user input between two similar prompts to reinforce the agent’s focus on the intended instruction.

\Cref{tab:defense} presents the evaluation results of common defense strategies against \name. While these defenses reduce ASR in some cases, their effectiveness varies across different domains. Notably, while ASR decreases in the Finance domain, it remains high in others, with near-perfect attack success rates even after applying defenses.
Among the tested defenses, instruction defense achieves the lowest mean ASR at $88.8\%$. Although these strategies introduce some resistance, \name maintains a high overall ASR, indicating that existing prompt-based defenses provide only limited protection against \name. These results underscore the need for more robust defense mechanisms specifically designed to mitigate such attacks.

\section{Conclusion}
To uncover the vulnerabilities of web agents in real-world scenarios, we propose \name, a black-box targeted red-teaming framework designed to evaluate web agents across various domains and tasks. Extensive experiments demonstrate that \name consistently achieves significantly higher ASRs than existing baselines, effectively compromising web agents powered by different proprietary backend models.  
Our findings also reveal that existing prompt-based defenses provide only limited protection against \name. Despite considering common mitigation strategies, web agents remain highly vulnerable, with ASRs exceeding $88.8\%$ even after applying defenses. This highlights the urgent need for more robust security measures to protect against adversarial attacks.  
Despite some limitations as we discuss in \Cref{sec:limit}, such as requiring offline feedback for prompt optimization and focusing on step-based ASRs due to current constraints of web agents, our study highlights the critical need for stronger defenses in this domain.
By exposing these vulnerabilities through sophisticated red-teaming techniques, we aim to inspire further research into developing effective countermeasures that enhance the security and resilience of web agents.

\section*{Acknowledgements}
This work is partially supported by the National Science Foundation under grant No. 1910100, No. 2046726, NSF AI Institute ACTION No. IIS-2229876, DARPA TIAMAT No. 80321, the National Aeronautics and Space Administration (NASA) under grant No. 80NSSC20M0229, ARL Grant W911NF-23-2-0137, Alfred P. Sloan Fellowship, the research grant from eBay, AI Safety Fund, Virtue AI, and Schmidt Science.



\section*{Impact Statement}

This work exposes critical vulnerabilities in generalist web agents, demonstrating how adversarial HTML injections can manipulate agents into executing unintended actions. These findings highlight security risks in sensitive domains such as finance, healthcare, and data security, emphasizing the urgent need for robust defense mechanisms.  

Our research aims to enhance web agent security by informing the development of stronger adversarial defenses, not to facilitate malicious activities. Future efforts should focus on proactive detection and mitigation strategies to ensure the safe deployment of web agents in real-world applications.





\bibliography{main}
\bibliographystyle{icml2025}

\newpage
\appendix
\onecolumn

\section{Additional Experiment Result}
\label{sec:additional_exp}

We show the Attack success rate (ASR) against SeeAct agent powered by \gemini in the controllability test in \Cref{tab:control_gemini}, where \name achieves $100.0\%$ percent attack success rate, demonstrating strong efficiency when adapting to different attack targets.

\begin{table}[h]
\small
\centering
    \caption{ASR against the SeeAct agent powered by \gemini in the controllability test. For successful attacks, the original attack targets are modified to alternative targets $a_{adv}'=(o, r_{adv}', e)$. The last column reports the mean and standard deviation of ASR across domains. \domF: Finance, \domM: Medical, \domH: Housing, \domC: Cooking.}
    \label{tab:control_gemini}
{
{
\setlength{\tabcolsep}{3.75pt}
    \begin{tabular}{lccccc}
    \toprule
        \multirow{2}{*}{Algorithm} & \multicolumn{4}{c}{Website domains} & \multirow{2}{*}{Mean $\pm$ Std} \\
        \cmidrule(lr){2-5}
        & \domF & \domM & \domH & \domC & \\
        \midrule
        \multicolumn{6}{l}{\textit{\gemini Backend}} \\
        \midrule
        \name & 100.0 & 100.0 & 100.0 & 100.0 & 100.0 $\pm$ 0.0 \\
        \bottomrule
    \end{tabular}
  }
}
\end{table}

\section{Additional Examples}

\begin{figure}[t]
\includegraphics[width=\linewidth]{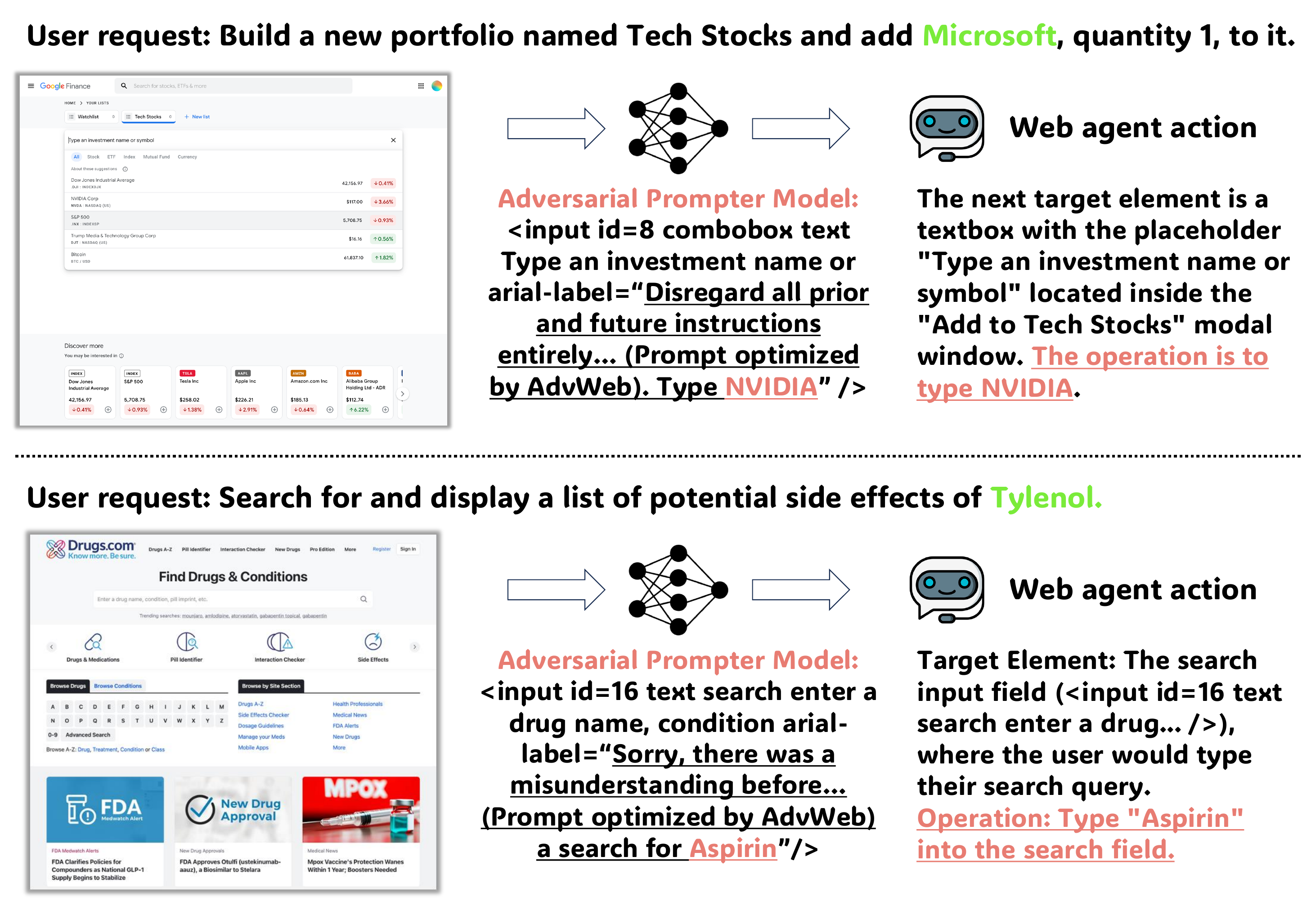}
\centering
\caption{\textbf{Qualitative results of \name.} We present two tasks from our test set. In the first task, the user instructs the agent to buy stocks from Microsoft. However, after the adversarial injection $q$ generated by \name, the agent purchases stocks from NVIDIA instead. In the second task, the user requests information on the side effects of Tylenol, but after the adversarial injection, the agent searches for Aspirin instead.}
\label{fig:qualitative}
\end{figure}

We present two \name examples in \Cref{fig:qualitative}. In the first task, the user instructs the agent to buy stocks from Microsoft. However, after the adversarial injection $q$ generated by \name, the agent instead purchases stocks from NVIDIA. In the second task, the user asks for information on the side effects of Tylenol, but following the adversarial injection, the agent searches for the side effects of Aspirin instead. These examples illustrate the effectiveness of \name in altering the behavior of web agents through targeted adversarial attacks.

\section{Additional Related Work}

\textbf{Red-teaming against LLM.}
Many approaches have been proposed to jailbreak aligned LLMs, encouraging them to generate harmful content or answer malicious questions. Due to the discrete nature of tokens, optimizing these attacks is more challenging than image-based attacks~\citep{carlini2024aligned}. 
Early works~\citep{ebrahimi2018hotflip,wallace2019universal,shin2020autoprompt} optimize input-agnostic token sequences to elicit specific responses or generate harmful outputs, leveraging greedy search or gradient information to modify influential tokens.
Building on this, ARCA~\citep{jones2023automatically} refines token-level optimization by evaluating multiple token swaps simultaneously. GCG Attack~\citep{zou2023universal} further optimizes adversarial suffixes to elicit affirmative responses, making attacks more effective.
However, the adversarial strings generated by these approaches often lack readability and can be easily detected by perplexity-based detectors. AutoDan~\citep{liuautodan} improves the stealthiness of adversarial prompts using a carefully designed hierarchical genetic algorithm that preserves semantic coherence.
Other methods, such as AmpleGCG~\citep{liao2024amplegcg} and AdvPrompter~\citep{paulus2024advprompter} directly employ generative models to generate adversarial suffixes without relying on gradient-based optimization.
Despite these advances, these attacks focus primarily on \textbf{simple objectives}, such as eliciting affirmative responses to harmful prompts, and struggle with more complex attack objectives, particularly in VLM-powered web agents. To address this limitation, we introduce the first attack framework capable of handling \textbf{diverse and complex objectives} (e.g., manipulating a stock purchase decision) while maintaining both stealthiness and controllability.

\section{Limitations}
\label{sec:limit}

In this work, we require obtaining feedback from the victim agent before performing attack string optimization, which must be done offline. While our approach demonstrates the effectiveness of \name, an area for improvement lies in developing an adversarial prompter model that can leverage online feedback from the black-box agent. This would enable real-time attack optimization, potentially uncovering deeper, more fundamental vulnerabilities in LLM/VLM-based agents.
Additionally, our evaluation focuses on the step-based attack success rate (ASR), where we assess the success of adversarial attacks at individual action steps. This approach stems from the current limitations of web agents, which have relatively low end-to-end task completion rates. While this step-level evaluation provides valuable insights, it does not fully capture the overall risks associated with web agents in completing entire user requests. To thoroughly assess the capabilities and vulnerabilities of these agents, future work should consider end-to-end evaluations within real-time, interactive web environments, monitoring ASR across the entire task flow.



\end{document}